\documentclass[a4paper]{article}

\usepackage{INTERSPEECH2020}
\usepackage{comment}
\usepackage{url}
\usepackage{graphicx}
\usepackage{subfig}
\usepackage{multicol}
\usepackage{makecell}
\usepackage{multirow}
\usepackage{xcolor}

\usepackage{soul}

\newcommand{\myparagraph}[1]{\vspace{.4em} \noindent \textbf{#1}\ }
\newcommand\blfootnote[1]{
  \begingroup
  \renewcommand\thefootnote{}\footnote{#1}
  \addtocounter{footnote}{-1}
  \endgroup
}

\title{Vector-Quantized Autoregressive Predictive Coding}

\name{Yu-An Chung, Hao Tang, James Glass}

\address{
  Computer Science and Artificial Intelligence Laboratory\\
  Massachusetts Institute of Technology\\
  Cambridge, MA 02139, USA}
\email{\{andyyuan,haotang,glass\}@mit.edu}

\begin{document}

\maketitle
\begin{abstract}
Autoregressive Predictive Coding~(APC), as a self-supervised objective, has enjoyed success in learning representations from large amounts of unlabeled data, and the learned representations are rich for many downstream tasks.
However, the connection between low self-supervised loss and strong performance in downstream tasks remains unclear.
In this work, we propose Vector-Quantized Autoregressive Predictive Coding~(VQ-APC), a novel model that produces quantized representations, allowing us to explicitly control the amount of information encoded in the representations.
By studying a sequence of increasingly limited models, we reveal the constituents of the learned representations.
In particular, we confirm the presence of information with probing tasks, while showing the \emph{absence} of information with mutual information, uncovering the model's preference in preserving speech information as its capacity becomes constrained.
We find that there exists a point where phonetic and speaker information are amplified to maximize a self-supervised objective.
As a byproduct, the learned codes for a particular model capacity correspond well to English phones.
\blfootnote{Code will be available at~\url{https://github.com/iamyuanchung/VQ-APC}.}
\end{abstract}

\section{Introduction}
Many high-level properties of speech, e.g., phonetic content and speaker characteristics, are not easily accessible%
\footnote{Following other recent works~\cite{oord2018representation,chung2019unsupervised,liu2020mockingjay}, in this study we use linear separability~(or separability with a shallow network) to define the accessibility of information for a downstream task.}
without sufficiently powerful transformations from the surface features such as audio waveforms and spectrograms.
Speech representation learning aims to search for a transformation from the surface features that better reveals these properties to downstream tasks.

Recently, self-supervised learning---a paradigm that treats the input itself or modifications of the input as learning targets---has obtained promising results for learning such transformations~\cite{oord2018representation,chung2019unsupervised,chorowski2019unsupervised,schneider2019wav2vec,pascual2019learning,baevski2020vq,baevski2020effectiveness,wang2020unsupervised,liu2020mockingjay,ling2020deep}.
These methods, mostly inspired by the techniques for pre-training NLP models~\cite{peters2018deep,radford2018improving,devlin2019bert}, learn speech representations by either inferring future information conditioned on historical audio~\cite{oord2018representation,chung2019unsupervised}, or predicting masked parts of input audio~\cite{wang2020unsupervised,liu2020mockingjay}.
The resulting representations, obtained without requiring any additional labeled data, are able to outperform the surface features across downstream tasks such as speech recognition, speech translation, speaker identification, and emotion recognition~\cite{chung2020generative,ravanelli2020multi}.

Autoregressive Predictive Coding~(APC)~\cite{chung2019unsupervised} is one of the recent self-supervised speech representation models.
APC defines a prediction task that trains an autoregressive neural model~(e.g., a unidirectional RNN) to predict a future frame considering the past context.
Although the learned representations contain highly accessible phonetic and speaker information, the reason why this seemingly unrelated self-supervised objective produces such a representation remains unclear.
In this work, we aim to provide an explanation, investigating the constituents that lead to low objective values, and connect them with the properties of the learned representations.

Our approach is to study the properties of the learned representation as we limit the model capacity. The models with limited capacity are forced to retain information to achieve maximal prediction, thereby allowing us to study the constituents of the task and the learned representations.
Several options are available to obtain a spectrum of models with different capacity, including reducing the number of layers, reducing the hidden layer size, or enforcing a bottleneck along the feed-forward process.
The impact of different numbers of hidden layers has been studied in prior work~\cite{chung2019unsupervised}.
Regardless, it is difficult to quantify the amount of information by changing the number of layers, changing the hidden layer size, or using low-rank matrices to enforce bottlenecks.
In this work, we study the use of vector quantization~(VQ), where the amount of information~(i.e., bits required to transmit the codebook and the sequence of codes) can be exactly quantified, to control the capacity of the models.

Recent studies on VQ for representation learning, mostly motivated by the discrete nature of phonetic units, attempt to show that enforcing the quantization leads to a better representation for acoustic unit discovery~\cite{van2017neural,harwath2020learning} and automatic speech recognition~\cite{baevski2020vq,liu2020towards}.
In contrast, our goal is not to discover the discrete units in speech.
We treat VQ as a general approach to limit the model capacity, and study its impact on the information encoded in the learned representations.

\section{Proposed Models}
\label{sec:proposed}

\subsection{A review on APC}
Given a sequence of acoustic feature vectors~$(x_{1}, x_{2}, ..., x_{t})$ as context, APC incorporates an autoregressive neural model~$g_{AR}$, e.g., a unidirectional RNN or a Transformer decoder~\cite{liu2018generating,radford2018improving}, to summarize the sequence for predicting a future frame~$x_{t + n}$ that is~$n$ steps ahead of~$x_{t}$.
Let~$y_{t}$ denote the prediction of~$g_{AR}$ at time~$t$.
In practice, for a speech utterance~$\mathbf{x} = (x_{1}, x_{2}, ..., x_{T})$, where~$T$ is the sequence length,~$g_{AR}$ is trained by minimizing the~$L1$ frame-wise loss between the predicted sequence~$(y_{1}, y_{2}, ..., y_{T - n})$ and the target sequence~$(x_{1 + n}, x_{2 + n}, ..., x_{T})$:
\begin{equation}
  \label{eq:apc-objective}
  \sum_{t = 1}^{T - n}|x_{t + n} - y_{t}|.
\end{equation}
Once~$g_{AR}$ is trained, one can extract features by taking its hidden representations, e.g., the last layer output, to replace the surface features as the new input to downstream models.

\begin{figure}[t]
  \centering
  \includegraphics[width=0.72\columnwidth]{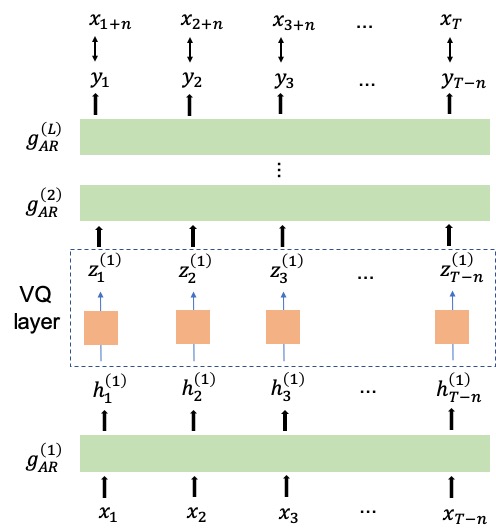}
\caption{A diagram of VQ-APC. $g_{AR}$ is an autoregressive model with~$L$ layers with~$g_{AR}^{(\ell)}$ denoting the~$\ell$-th layer and~$h_{t}^{(\ell)}$ denoting the output vector of~$g_{AR}^{(\ell)}$ at time~$t$. The figure is an example of inserting a VQ layer~(area inside the dashed box) between the first and second layers~$g_{AR}^{(1)}$ and~$g_{AR}^{(2)}$. The orange block represents the code variable lookup process that replaces~$h_{t}^{(\ell)}$ by~$z_{t}^{(\ell)}$, where~$z_{t}^{(\ell)}$ is one of the elements in a codebook. The quantized hidden vectors are fed to the next layer and the feed-forward process continues. The training objective is the same as APC: to minimize the~$L1$ loss between the predicted frame~$y_{t}$ and the target future frame~$x_{t + n}$.}
  \label{fig:vq-apc}
\end{figure}

\subsection{VQ-APC}
\label{sec:vqapc}
Figure~\ref{fig:vq-apc} illustrates the VQ-APC architecture, which is based upon APC with additional quantization layer(s).
Assume~$g_{AR}$ has~$L$ layers.
Let~$g_{AR}^{(\ell)}$ denote the~$\ell$-th layer of~$g_{AR}$.
After feeding~$\mathbf{x} = (x_{1}, x_{2}, ..., x_{T})$ to~$g_{AR}$, each~$g_{AR}^{(\ell)}$ will produce a sequence of hidden vectors~$\mathbf{h}^{(\ell)} = (h_{1}^{(\ell)}, h_{2}^{(\ell)}, ..., h_{T}^{(\ell)})$.
Next, we add a vector quantization~(VQ) layer~\cite{van2017neural} that replaces~$h_{t}^{(\ell)}$ by~$z_{t}^{(\ell)}$, where~$z_{t}^{(\ell)}$ is one of the~$V$ elements in a codebook~$\{c_1, \dots, c_{V}\}$.
We then pass the resulting hidden vectors $\mathbf{z}^{(\ell)} = (z_{1}^{(\ell)}, z_{2}^{(\ell)}, \dots, z_{T}^{(\ell)})$ to the next layer~$g_{AR}^{(\ell + 1)}$ and continue the feed-forward process.

We use the Gumbel-Softmax with the straight-through estimator~\cite{jang2017categorical} for selecting discrete codebook variables in a fully differentiable way.
Specifically, we apply a linear layer to map~$h_{t}^{(\ell)}$ to a vector~$r\in \mathbb{R}^{V}$.
At test time, we simply choose the largest index in~$r$.
At training time, the probability~$p_{i}$ of selecting the~$i$-th code variable~$c_{i}$ is computed as follows:
\begin{equation}
  p_{i} = \frac{e^{(r_{i} + v_{i}) / \tau}}{\sum_{j = 1}^{V}e^{(r_{j} + v_{j}) / \tau}},
  \label{eq:gumbel}
\end{equation}
where~$v = -\ln(-\ln(u)))\in \mathbb{R}^{V}$ and~$u$ is uniformly sampled from~$\mu (0, 1)$.
$\tau$ controls how close the approximation is to argmax.
During the forward pass, the argmax code~$c_{k}$ where~$k = \mathrm{argmax}_{i}p_{i}$ is chosen; during the backward pass, the true gradients of the Gumbel-Softmax outputs are used.
The training objective is the same as APC~(Equation~\ref{eq:apc-objective}).

The codebook size and the code dimension of a VQ layer control the amount of information from the previous layer flowing to the next, and changing either of these two factors allows us to explicitly control the capacity of the models.
As we limit the model capacity, the model is forced to retain information to achieve maximal prediction.
By studying a sequence of increasingly limited models, we are able to reveal the constituents of the prediction task and the learned representations.


\section{Experiments}
\label{sec:exps}
We conduct experiments to study the properties of the learned representations and their connection to the self-supervised objective.
Since VQ layers are known to significantly disrupt model training, we first examine where VQ layer(s) should be inserted.
Next, by using phonetic and speaker classification as probing tasks, we show the model's preference in preserving speech information as its capacity becomes constrained.
We then visualize the learned VQ codes to show the presence and absence of phonetic information and the correspondence between codes and phones.
Finally, we compare VQ-APC with other self-supervised speech representation models.

\subsection{Setup}
\myparagraph{Training of self-supervised models.}
All self-supervised models, including the VQ-APC variants and other models to be compared, are trained on the clean 360-hour subset of LibriSpeech~\cite{panayotov2015librispeech}.
We use~80-dimensional log Mel spectrograms~(normalized to zero mean and unit variance per speaker) as input features, that is,~$x_t\in \mathbb{R}^{80}$, and train all models for~100 epochs using Adam~\cite{kingma2015adam} with a batch size of~32 and an initial learning rate of~$10^{-3}$.

\myparagraph{Probing tasks.}
We consider phonetic and speaker classification for measuring the accessibility of the phonetic and speaker information contained in the representations, respectively.
Both experiments are carried out on the Wall Street Journal corpus~(WSJ)~\cite{paul1992design}.
For phonetic classification, the goal is to correctly classify each frame in an utterance into one of the~42 phones.
The phone alignments are generated with a speaker adapted GMM-HMM model.
We follow the standard split of WSJ, using~90\% of {\tt si284} for training, the rest~10\% for validation, and reporting phone error rate on {\tt dev93}.
For speaker classification, the goal is to correctly predict the speaker identity of an utterance.
We follow~\cite{chung2020generative} and consider a 259-class classification task where each class corresponds to an unique speaker, using~80\% of {\tt si284} for training, the other~10\% for validation, and reporting classification error rate on the rest~10\%.
We note that speaker classification is not a typical task for WSJ, and only serves as a sanity check for the presence of speaker information~(and its potentially correlated channel information)~\cite{oord2018representation,liu2020mockingjay}.
The classifier for both tasks is a linear logistic regression that takes the features extracted from the self-supervised models as input.
For speaker classification, the features from the same utterance are averaged before being fed to the classifier.
All self-supervised models are kept frozen when training the linear classifier.
All reported numbers are an average of~5 runs, of which variances are negligibly small and not included.

\begin{table}[t]
  \caption{Phonetic classification results of VQ-APC with different VQ configurations. The layers where VQ is inserted are denoted as a set, and~$\emptyset$ is equivalent to the regular APC. We compare both the hidden vectors~$\mathbf{h}^{(\ell)}$ and their corresponding VQ codes~$\mathbf{z}^{(\ell)}$~(when applicable) for~$\ell = 1, 2, 3$ as extracted features. Training loss on LibriSpeech is also reported. The lowest phone error rate is highlighted in bold.}
  \label{tab:vq-config}
  \centering
  \resizebox{\columnwidth}{!}{
    \begin{tabular}{lccccccc}
      \toprule
      \multirow{2}{*}{\makecell{VQ\\config.}}  &  \multirow{2}{*}{\makecell{Train\\loss}}  &  \multicolumn{6}{c}{Phone error rate} \\
      \cmidrule(lr){3-8}
                          &  &  $\mathbf{h}^{(1)}$  &  $\mathbf{h}^{(2}$  &  $\mathbf{h}^{(3)}$  &  $\mathbf{z}^{(1)}$  &  $\mathbf{z}^{(2)}$  &  $\mathbf{z}^{(3)}$  \\
      \midrule
      \midrule
      $\emptyset$         &  0.68  &  46.5  &  38.6  &  33.3  &  $-$   &  $-$   &  $-$  \\
      \midrule
      $\{1\}$             &  0.70  &  43.0  &  37.5  &  32.3  &  43.4  &  $-$   &  $-$  \\
      $\{2\}$             &  0.72  &  45.8  &  35.4  &  31.5  &  $-$   &  36.0  &  $-$  \\
      $\{3\}$             &  0.73  &  46.4  &  35.7  &  {\bf 30.5}  &  $-$   &  $-$   &  30.8 \\
      $\{1, 2\}$          &  1.22  &  75.0  &  72.3  &  70.7  &  74.8  &  72.6  &  $-$  \\
      $\{1, 3\}$          &  0.83  &  59.9  &  54.1  &  51.0  &  61.2  &  $-$   &  51.4 \\
      $\{2, 3\}$          &  0.87  &  63.1  &  58.7  &  54.7  &  $-$   &  59.9  &  55.2 \\
      $\{1, 2, 3\}$       &  1.21  &  75.3  &  74.1  &  68.5  &  75.4  &  75.2  &  67.8 \\
    \bottomrule
    \end{tabular}
  }
\end{table}

\subsection{Preliminary VQ experiments}
We first explore several potential places to insert VQ layers.
For all VQ-APC variants in our experiments, the autoregressive model~$g_{AR}$ is a~3-layer unidirectional GRU with~512 hidden units, and the target frame in the future,~$n$, is set to~5 when training~(Equation~\ref{eq:apc-objective}) on LibriSpeech.
Whenever a VQ layer is added, the embedding dimension of each code is~512, and~$\tau$ for the Gumbel-Softmax straight-through estimator~(Equation~\ref{eq:gumbel}) is a fixed value of~0.1 throughout training.

Table~\ref{tab:vq-config} presents the phonetic classification results of adding VQ layers to different layers in~$g_{AR}$.
In the ``VQ config.'' column, the numbers inside the parenthesis denote the layers we insert a VQ layer.
For example,~$\{1\}$ means that we only add VQ layer after~$g_{AR}^{(1)}$.
$\emptyset$ denotes the case where no VQ layer is applied, equivalent to the regular APC.
The codebook size here is~128.
We try using both the hidden vectors~$\mathbf{h}^{(\ell)}$ and their quantized codes~$\mathbf{z}^{(\ell)}$~(when applicable) for~$\ell = 1, 2, 3$ as extracted features.
We also include the final VQ-APC training loss on the LibriSpeech 360-hour subset after~100 epochs~(not the downstream linear classifier's training loss on WSJ).

\myparagraph{Quantizing one layer.}
As indicated by the training loss, we see that the bottleneck imposed by the VQ layer indeed handicaps the models' ability to predict the future, as~$\{1\}, \{2\}$, and~$\{3\}$ all have higher training loss than~$\emptyset$.
In terms of phone error rate, regardless of where VQ is inserted, we see improvement over the APC representations.
Inserting VQ at the third layer leads to the most improvement, from~33.3 to~30.5.
The quantized codes~$\mathbf{z}^{(\ell)}$, when applicable, could also be used as extracted features, which perform similarly to their corresponding pre-quantized representations.
For example, in~$\{3\}$,~$\mathbf{z}^{(3)}$~(30.8) is close to~$\mathbf{h}^{(3)}$~(30.5).


\begin{figure}[t]
  \centering
  \includegraphics[width=0.8\columnwidth]{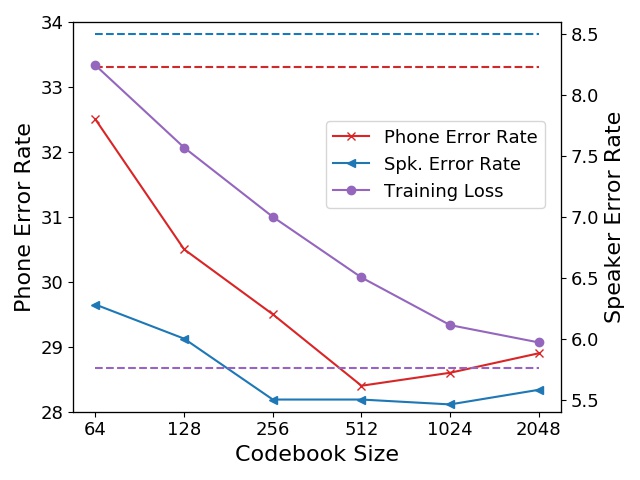}
  \caption{Training loss~(purple), phonetic classification~(red), and speaker classification~(blue) results of VQ-APC with VQ configuration~$\{3\}$ using varying codebook sizes. The x-axis is the codebook size, decreasing from~2048 to~64, and the y-axis on the left is the corresponding phone error rate and on the right the speaker~(spk.) error rate. For clarity, the vertical axis for training loss is not displayed. The three dash horizontal lines show the corresponding results of a regular APC, i.e.,~$\emptyset$.}
  \label{fig:varying-codebook-size}
\end{figure}

\myparagraph{Quantizing multiple layers.}
We find that our VQ-APC models with multiple VQ layers have trouble fitting the training set.
Their representations are also much worse than the regular APC on phonetic classification.
One potential remedy is to enable VQ with a schedule~\cite{harwath2020learning}, but is beyond the scope of the paper.

\subsection{The constituents of the learned representations}
Experiments so far suggest that the phonetic information is still present~(if not better) after using VQ.
For the rest of the paper, we will focus on the case where VQ is inserted at the third layer, i.e., the case of $\{3\}$.
To study the constituents of the learned representations, we train a series of increasingly limited VQ-APC models by decreasing the codebook size from~2048 to~64 while fixing the code dimension to~512.
As the codebook size becomes smaller, the model is forced to choose what information to encode and what to discard, thus revealing the constituents of the learned representations.
We show the training losses of these models at convergence and the respective phone and speaker error rates in Figure~\ref{fig:varying-codebook-size}.
The dashed lines are the training loss, phone error rate, and speaker error rate of a regular APC model.

First, the training loss~(purple curve) increases as expected, showing worse fit on the training set as we limit the codebook size.
Note that in theory, when the codebook size goes to infinity, we recover the regular APC.
The phone error rate~(red curve) obtains a minimum at codebook size~512, and starts to worsen with smaller codebook size.
The sharp degradation in phone error rate suggests that the model discards certain phonetic information to achieve maximal self-supervised objective.

The speaker error rate~(blue curve), on the contrary, does not change by much as we limit the codebook size.
This shows that the speaker information~(and its potentially correlated channel information) is mostly retained.
Given the sharp degradation in phone error rate, we can conclude that the model prefers to retain speaker information over phonetic information to achieve maximal future prediction.
The preference of information can potentially stem from the use of GRUs, the VQ configuration, and the self-supervised, future prediction objective.
More analyses are needed to disentangle the among these causes.

On the other hand, when the codebook size becomes large, the model falls back to regular APC and might suffer from overfitting~\cite{chung2020improved}, paying unnecessary attention to the spectral details that does not generalize for predicting future frames.
Finally, even with a codebook size of~64, we still see gains over regular APC, showing the strong performance of VQ-APC in representation learning.


\begin{figure*}
  \centering
  \begin{multicols}{3}
    \includegraphics[width=0.33\textwidth]{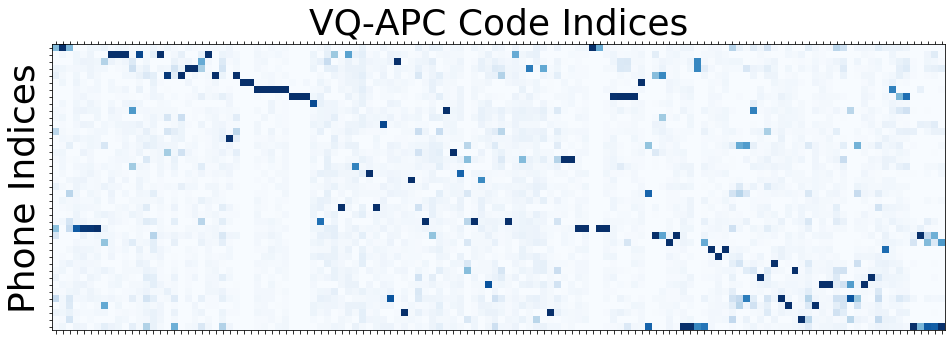}
    \includegraphics[width=0.33\textwidth]{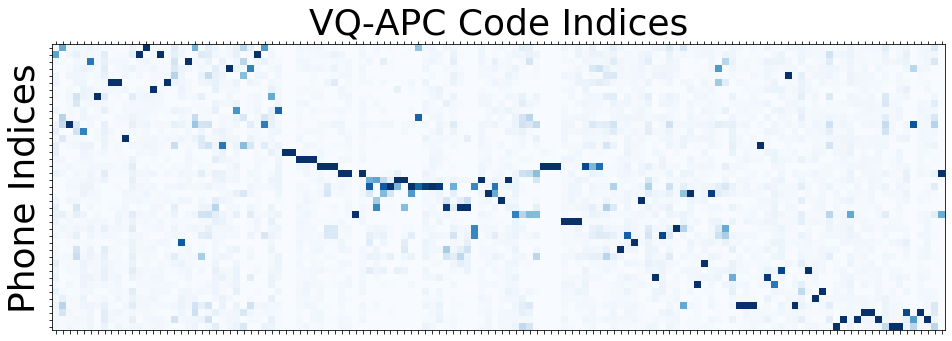}
    \includegraphics[width=0.33\textwidth]{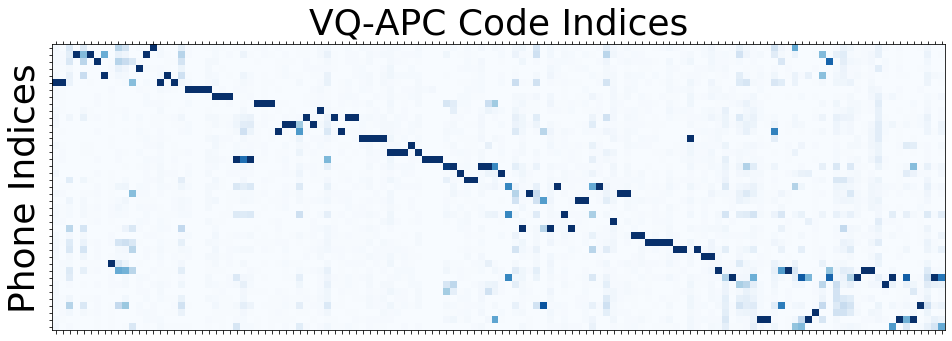}
  \end{multicols}
  \caption{From left to right, visualizations of the conditional probability matrix~$P(\mathrm{phone}|\mathrm{code})$ for configurations~$\{1\}, \{2\}$, and~$\{3\}$ with~128 codes, respectively. Each sub-figure is a~$42\times 128$ conditional probability matrix, where each row and column correspond to a phone and code index, respectively. Color scaling is saturated at probability~0.5 for better visualization.}
  \label{fig:phn-code-cooccurrence}
\end{figure*}

\subsection{Visualizations of learned codes}
To better measure the correspondence between the learned VQ codes and English phones, we compute co-occurrence statistics~(at the frame level) across the 360-hour subset of LibriSpeech, the dataset we use to train the VQ-APC models.
We compare three settings, $\{1\}$, $\{2\}$, and $\{3\}$ with a codebook size of~128.
The conditional probability $P(\text{phone}|\text{code})$, as shown in Figure~\ref{fig:phn-code-cooccurrence}, are estimated based on the co-occurrence statistics, i.e., via maximum likelihood.
In each sub-figure, the rows and columns are ordered via spectral co-clustering with~15 clusters to group together phones that share similar sets of codes, and a diagonal segment would imply a high correspondence between phones and codes.
Note that the phone labels of LibriSpeech are only used for analysis and never seen during training.

From Figure~\ref{fig:phn-code-cooccurrence}, we see that the correspondence between phones and VQ codes is stronger when quantized at higher layers, and is especially strong for~$\{3\}$.
Recall that probing tasks are useful for showing the presence of certain information, but have difficulty showing the absence of it.
In contrast, given the co-occurrence statistics, mutual information can be estimated to support the absence of information.
The normalized mutual information are~0.167,~0.285, and~0.406 for $\{1\}$, $\{2\}$, and $\{3\}$, respectively.
In other words, not only can we conclude that the learned representations in $\{3\}$ contain phonetic information, we can also readily conclude that $\{1\}$ and $\{2\}$ contain much less information for certain phones.

\subsection{A comparison with other self-supervised models}
Finally, we compare VQ-APC with other self-supervised speech representation models, including Contrastive Predictive Coding~(CPC)~\cite{oord2018representation}, Bidirectional Masked Reconstruction~\cite{wang2020unsupervised}, Mockingjay~\cite{liu2020mockingjay}, and Multi-Target APC~\cite{chung2020improved}.
We briefly review these methods below, and show their results on phonetic and speaker classification in Table~\ref{tab:ss-compare}.
To stay as close to the original implementation as possible, we do not separate the discussion of models, such as the use RNNs or Transformers, and the self-supervised objectives.

CPC and APC share a similar methodology as both use an autoregressive model
to learn representations through conditioning on past frames to predict information about a future frame.
Their difference is that while APC tries to directly predict the future frame via regression, CPC aims to learn representations containing information that most discriminates the future frame from a set of negative samples.
We mainly follow the original paper~\cite{oord2018representation} for implementing CPC with some modifications described in~\cite{chung2019unsupervised}.
These modifications are meant to minimize the architectural differences between APC and CPC while maintaining their training objectives.

Multi-Target APC~(MT-APC) is an extension of APC.
It incorporates an auxiliary objective serving as a regularizer to improve the generalization of the main future prediction task.
The exact same setup described in~\cite{chung2020improved} is used in our experiments.

Different from CPC and APC that are based on the idea of future prediction,
Bidirectional Masked Reconstruction~(BMR) and Mockingjay are under the category of {\it masked prediction}.
Inspired by the masked language modeling technique from BERT~\cite{devlin2019bert}, BMR and Mockingjay mask parts of the input signals, and predict them through conditioning on both past and future contexts with a bidirectional RNN and Transformer encoder~\cite{vaswani2017attention}, respectively.
For experiments, we mainly follow the implementations in the original papers~\cite{wang2020unsupervised,liu2020mockingjay}, except that the number of layers are reduced to match ours to minimize the architectural differences.

\begin{table}[htbp]
  \caption{Phonetic and speaker classification results of different self-supervised speech representation models. All features are fed to a linear logistic repression. For log Mel, we also include the results of using a 1- and 3-layer multi-layer perceptron, denoted as MLP-1 and MLP-3, respectively. We also note the neural architectures used by each model.}
  \label{tab:ss-compare}
  \centering
    \resizebox{\columnwidth}{!}{
  \begin{tabular}{lccc}
    \toprule
    Model             &  Network   &  \makecell{Phone\\error rate}   &  \makecell{Speaker\\error rate}\\
    \midrule
    \midrule
    log Mel           &     $-$    &  50.3  &  17.6 \\
    log Mel + MLP-1   &     $-$    &  43.1  &  12.3 \\
    log Mel + MLP-3   &     $-$    &  41.2  &  11.9 \\
    \midrule
    CPC~\cite{oord2018representation}          &  3-layer uni-GRU      &  34.1  &  9.7 \\
    APC~\cite{chung2019unsupervised}           &  3-layer uni-GRU      &  33.3  &  8.5 \\
    MT-APC~\cite{chung2020improved}            &  3-layer uni-GRU      &  30.5  &  7.3 \\
    VQ-APC~(ours)                              &  3-layer uni-GRU      &  28.4  &  5.5 \\
    \midrule
    BMR~\cite{wang2020unsupervised}            &  3-layer bi-GRU       &  32.4  &  6.2 \\
    Mockingjay~\cite{liu2020mockingjay}        &  3-layer Transformer  &  30.8  &  5.1 \\
    \bottomrule
  \end{tabular}
  }
\end{table}

On phonetic classification, we see that VQ-APC~(28.4) improves over APC~(33.3) and MT-APC~(30.5), demonstrating the effectiveness of VQ layers.
It also outperforms other self-supervised models despite using the same~(vs.\ CPC) or smaller~(vs.\ BMR and Mockingjay) network.
On speaker classification, VQ-APC~(5.5) again improves over the other two APC models~(8.5 and~7.3), and is on par with the best model~(Mockingjay, 5.1).
On both tasks, all self-supervised models outperform log Mel regardless of the type of classifier it uses.

\section{Conclusions}
\label{sec:cons}
We have demonstrated that incorporating vector quantization~(VQ) layers into an Autoregressive Predictive Coding model imposes a bottleneck, forcing the model to learn better representations.
Extensive experiments have been conducted to compare different VQ configurations, to study the effect of varying codebook sizes, and to compare with other self-supervised speech representation models.
We show evidence for the presence and absence of phonetic and speaker information in the learned representations, and also show the model's preference in retaining information when the model capacity is limited, in the hope to bridge the connection between the self-supervised objective and the properties of the learned representations.
When the phonetic information is present, the learned VQ codes also correspond well with English phones.

\bibliographystyle{IEEEtran}
\bibliography{mybib}

\end{document}